\documentclass[aps,pra,twocolumn,groupedaddress]{revtex4-1}

\usepackage{graphicx}
\usepackage{braket}
\usepackage{amssymb}

\newcommand{\refeq}[1]{Eq.~(\ref{#1})}
\newcommand{\refsec}[1]{Section~\ref{#1}}
\newcommand{\reffig}[1]{Figure~\ref{#1}}

\begin{document}

\title{Testing the Robustness of Robust Phase Estimation}

\author{Adam M. Meier}
\author{Karl A. Burkhardt}
\author{Brian J. McMahon}
\author{Creston D. Herold}
\email[]{Creston.Herold@gtri.gatech.edu}
\affiliation{Georgia Tech Research Institute, Atlanta, GA 30332, USA}

\date{\today}

\begin{abstract}
The Robust Phase Estimation (RPE) protocol was designed to be an efficient and
robust way to calibrate quantum operations. The robustness of RPE refers
to its ability to estimate a single parameter, usually gate amplitude, even when
other parameters are poorly calibrated or when the gate experiences significant
errors. Here we demonstrate the robustness of RPE to
errors that affect initialization, measurement, and gates. In each
case, the error threshold at which RPE begins to fail matches quantitatively with
theoretical bounds. We conclude that RPE is an effective and reliable tool for
calibration of one-qubit rotations and that it is particularly useful for
automated calibration routines and sensor tasks.
\end{abstract}

\maketitle

\section{Introduction}
The standard, circuit model of quantum computation requires discrete, unitary
operations (gates). However, physical implementation of these gates is not
naturally discrete. For one-qubit gates it is
useful to picture a unitary operation as a specific rotation of the Bloch
sphere, with an axis and an angle of rotation that can be parametrized by three
real numbers. Careful calibration of these parameters is necessary in order to
compute accurately using the gate \cite{Wright2019,Erhard2019,Kelly2018}.

When a single unitary operation is concatenated into a sequence of several
identical operations, the resulting angle of rotation is linear in the number
gates. Measurements of the quantum state after such a sequence can cause small
changes in the angle of the individual gate to appear quadratically as changes in the
resulting projection probabilities. As a result, the error of a computation can
scale as the square both of the angular error on each operation and of the number
of operations.  Because of this scaling, small errors in calibrating operation
rotation angles using simple calibration experiments can still lead to
significant error in the context of longer gate sequences.

In its original form, the Robust Phase Estimation (RPE) protocol was proposed as
a practical method for estimating the axis and angle of a single qubit
rotation. The protocol was originally developed by \citet{Kimmel2015} and tested
in experiments by \citet{Rudinger2017}. The robustness of RPE refers to its
ability to estimate these two parameters even when other parameters of the experiment are poorly
calibrated or prone to stochastic errors. In this paper we
describe three separate experiments that demonstrate the robustness of RPE to
injected errors chosen to represent common experimental error sources.

The RPE protocol consists of performing gate sequences of several different lengths; in
each sequence the same gate is repeated, so that the final rotation angle compounds linearly.
The gate sequences themselves are very similar to the standard ``Rabi flopping''
procedure, in which the rotation interaction is applied continuously for varying
durations before qubit measurement. The result of Rabi flopping is ideally a sinusoidal
dependence of final state population on interaction duration, from which one can extract a “Rabi
frequency” that can in turn be used to calculate the rotation angle realized in
any finite duration. The primary differences between Rabi flopping and RPE are
that RPE repeats an operation with a fixed duration, instead of continuously
varying the duration of a single operation, and that the structure and analysis
procedure of RPE are more complex. 

In this work, we focus only on using RPE to estimate the angle of the rotation,
called $\theta$ in this work, and not on estimating the rotation axis (called
$\theta$ in the original formulation). From a practical standpoint, RPE can
determine the rotation angle of a fixed gate duration much more efficiently than
Rabi flopping, where efficiency is measured by the number of gate applications
required to achieve a given precision in the estimation of the angle.

RPE succeeds or fails in a binary
fashion; the estimate it produces is either correct (within a pre-determined
confidence region) or it is incorrect, sometimes by a large angle. The protocol
can fail at each sequence length (independently) due to shot noise or to errors
in the implementation of the quantum operations. When each sequence is repeated
sufficiently many times to remove shot noise concerns and the operation errors
are small enough, the protocol always succeeds in producing an estimate of the rotation
angle within its guaranteed confidence region. The robustness of RPE is defined
by this feature that the protocol reliably succeeds even when these errors are
significant.

Rabi flopping experiments are useful for diagnostic experiments, where
protocol efficiency is less important and an experimenter can identify and
process unexpected results.  In contrast, the robustness and efficiency of RPE
make it ideal for automated calibration of quantum systems or for automated sensor
experiments where there is no human oversight. Gate set tomography (GST) is
another useful calibration protocol for diagnosing quantum computing
experiments. Where the RPE protocol attempts to estimate one or two parameters
of a gate while ignoring imperfections in others, GST attempts to fully
characterize all parameters of the gate. Ref.~\cite{Rudinger2017} clearly
examines the differences in cost and utility for the two protocols and finds
that RPE is significantly more efficient for single parameter calibrations.

The remainder of this paper is organized as follows. In \refsec{sec:protocol},
we summarize our implementation of the RPE protocol. Our experimental apparatus
is described in \refsec{sec:experiment}. In \refsec{sec:errors}, we describe the
errors against which we test the robustness of RPE. The results of these tests
follow in \refsec{sec:results}. Finally, we comment in \refsec{sec:conclusion} on
the wide range of parameter estimation experiments which could benefit from RPE.

\section{\label{sec:protocol}RPE Protocol}
In this section, we provide a practical description of the RPE protocol as
applied in our experiments.  Consider calibrating the rotation angle $\theta$ of
a qubit operation described by
\begin{equation}
	\label{Y-theta}
	Y_{\theta} = \cos(\theta/2) \mathbb{I} - i \sin(\theta/2) \sigma_Y 
\end{equation}
where $\mathbb{I}$ is the identity matrix and $\sigma_Y$ is the Pauli $Y$-matrix.

We perform two sets of experiments, where the operation of interest is repeated
$\{n\}$ times for various values of $n$, and the sets are differentiated by the
initial qubit state. The experiments all terminate with a single, projective
measurement. We repeat each experiment $M_n$ times (subsequently
referred to as ``samples'') and record the number of times we observe the +1
eigenstate. The observed number of the +1 results for the two sets of
experiments are denoted by $x_n$ and $y_n$, and their values in the limit of
large $M_n$ can be predicted using the following formulas:
\begin{eqnarray}
	x_n = M_n \left|\braket{1 | Y_{\theta}^n | 0} \right|^2
	\qquad
	y_n = M_n \left|\braket{1 | Y_{\theta}^n | +} \right|^2 ,
\end{eqnarray}
where the state $\ket{+}$  is equal to $(\ket{0}+\ket{1})/\sqrt{2}$ and
$\{\ket{0},\ket{1}\}$ are the $Z$-eigenstates of the qubit. In our experiment we
always initially prepare $\ket{0}$. The state $\ket{+}=Y_{\pi/2}\ket{0}$ is
generated from the initial state with a $Y_{\pi/2}$ operation.

In order to account for non-idealities in the experiments, we follow the
convention of Ref.~\cite{Kimmel2015} and write the expected probability of
observing the +1 eigenstate in the experiments corresponding to $x_n$ and $y_n$
as
\begin{eqnarray}
	\label{eq:additive}
	p_x(n) &=& \frac{1-\cos(n\theta)}{2} \pm \delta_x(n) \nonumber
	\\
	p_y(n) &=& \frac{1+\sin(n\theta)}{2} \pm \delta_y(n) ,
\end{eqnarray}
where we have introduced additive error terms $\delta_{x,y}$ which may be
$n$-dependent. When the $\delta_{x,y}$ terms are zero, we can estimate the total
rotation $n\theta$, modulo $2\pi$, by noting that
\begin{equation}
	\tan(n\theta) = \frac{2 p_y(n) -1}{1-2 p_x(n)},
\end{equation}
and using $x_n/M_n$ and $y_n/M_n$ as estimates of $p_{x,y}(n)$ gives
\begin{equation}
	n\theta = \tan^{-1}\left(\frac{y_n-M_n/2}{M_n/2-x_n}\right) ,
\end{equation}
We note that the inability to discriminate $n\theta = 0$ from $2\pi$ can
typically be remedied in practice, by repeating the protocol with a gate using
the same physics but a smaller intended angle.
 
When the error terms $\delta_{x,y}$ are non-zero, our estimate of $n\theta$ can
be biased. Additionally, if we collect too few samples, shot noise, which is
proportional to $1/\sqrt{M_n}$, can lead to erroneous results. These
considerations are described in great depth in Ref. \cite{Kimmel2015}.

The RPE protocol achieves high efficiency by specifying that only certain values
of $n$ (gate repetitions) need to be performed, namely $n = 2^{j-1}$ where
$j \in \{1,2,3,\ldots,L\}$ and $L$ is chosen as a function of the desired precision
of the estimate. The $j$th measurement restricts the estimate of $\theta$:
\begin{equation}
	\label{eq:range}
	\theta - \theta_{\mathrm{actual}} 
	\in \left( -\frac{\pi}{2^j}, \frac{\pi}{2^j}\right].
\end{equation}
Each successive doubling of the repetition number cuts the range of the angle
estimate bound in half. The supplemental information to \cite{Rudinger2017}
provides pseudocode for determining $\theta$ including enforcement of the range
restriction from preceding steps. Ref. \cite{Kimmel2015} also provides the
projected probability of protocol success conditioned only on statistical
errors. Here success is defined to be an application of the protocol in which
the true value of $\theta$ lies within the bounds output by the protocol. With
the range restriction provided in \refeq{eq:range}, Kimmel et al.~proved in
\cite{Kimmel2015} that RPE is robust in the presence of errors and will
succeed provided the total additive error is bounded by
$|\delta_{x,y}(n)| < \delta_{\mathrm{bound}} \equiv 1/\sqrt{8}$ for all
repetitions $\{n\}$.

In practice, it is likely to be difficult to estimate the additive error a
priori. For example, without knowing the angle of the gate under inspection, the
experimenter will not know the expected final state of the length $n$ sequence
and cannot estimate the impact of asymmetric measurement errors. Even more
difficult is the case of gate errors associated with colored noise, where the
apparent error magnitude scales non-linearly with the number of gates in the
sequence.  Errors that are mutually uncorrelated, such as measurement errors or
errors due to decay events, are simpler to bound as a function of sequence
length, and we focus on them in this manuscript.

It is important to note that as additive error increases, more samples are
required for RPE to succeed, and the required $M_n$ diverges at the $1/\sqrt{8}$
bound, as described in \cite{Kimmel2015}. A prescription for choosing $M_n$ as a
function of the predicted additive error and sequence length is provided
therein. Furthermore, because each sequence length constrains the next estimate
of $\theta$, the probability that RPE fails is the sum of the probability of
failing at each step.

In this paper, we demonstrate experimentally that, for three error sources that
are expected to behave additively, the RPE protocol is as robust as promised.
This is evaluated by observing the failure rate of RPE for the gate
$Y_{\pi/2}$ as a function of injected (artificially introduced) error sources.

\section{\label{sec:experiment}Experimental System}
\begin{figure}
	\includegraphics[width=0.48\textwidth,]{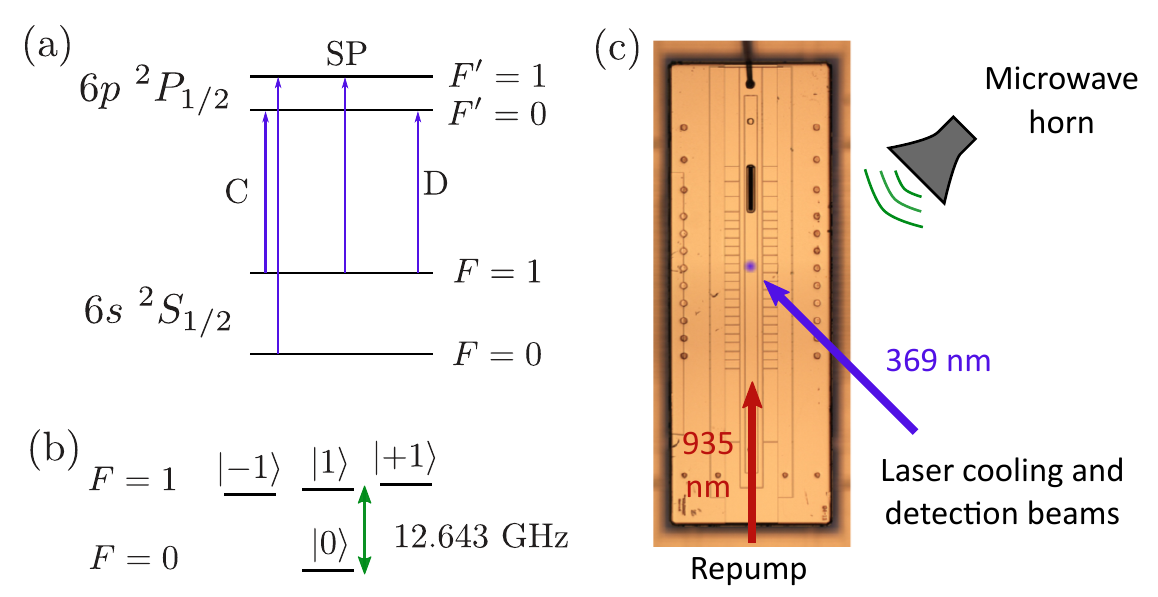}
	\caption{\label{fig:experiment}
	(a) $^{171}\mathrm{Yb}^+$ energy level diagram. The required 369 nm laser
	transitions are shown for Doppler cooling (C), initial state preparation
	(SP), and detection (D).
	(b) State labels in $^2S_{1/2}$ hyperfine manifold. The qubit levels are
	$\ket{0}$ and $\ket{1}$.
	(c) System overview. Laser beam geometry and microwave horn for qubit
	operations are overlaid on an image of the trap chip.
	}
\end{figure}

All experiments described herein are performed on a single
$^{171}\mathrm{Yb}^+$ ion trapped above a GTRI-Honeywell BGA trap
\cite{Guise2015}. The relevant portion of the atomic energy level diagram is
shown in Figure \ref{fig:experiment}a. Doppler cooling, state preparation, and
state detection are all performed with the same 369 nm beam through
electro-optic modulation, similar to \cite{Olmschenk2007a}. This beam is
switched on and off with an acousto-optic modulator (AOM). The qubit states
are chosen from the ground-state hyperfine manifold (\reffig{fig:experiment}b)
to be $\ket{0} \equiv \ket{F=0,m_F=0}$ and $\ket{1} \equiv \ket{F=1,m_F=0}$. We
label the auxiliary states $\ket{\pm1} \equiv \ket{F=1,m_F=\pm1}$.

One-qubit rotations are driven with microwaves emitted from a horn outside the
vacuum chamber as depicted in Figure \ref{fig:experiment}c. We control the
frequency, amplitude, and phase of the microwave drive by mixing an
approximately 200 MHz DDS signal with a local oscillator at $12.442\,812$~GHz.
The upper sideband of this mixing is resonant with the transition from
$\ket{0}$ to $\ket{1}$. The mixer output is amplified to a maximum of
approximately 1.6~W, high enough to achieve a minimum $\pi/2$-rotation time of
$T_{\pi/2} \approx 4~\mu$s. However, for all the reported RPE measurements, we
reduce the amplitude and use $T_{\pi/2} = 10$ to 30~$\mu$s to reduce amplitude
errors caused by the thermal duty cycle of the microwave amplifier chain.

While RPE can be used to calibrate any rotation, we choose $Y_{\pi/2}$ (i.e.
$\theta=\pi/2$ in \refeq{Y-theta}) for our base gate. This means that we prepare
$\ket{+}=(\ket{0}+\ket{1})/\sqrt{2}$ with the same gate we repeat for the RPE
sequence. Prior to each measurement set, we use an RPE sequence with no intentionally
injected errors to calibrate the $\pi/2$ operation for a fixed $T_{\pi/2}$ by
adjusting the microwave amplitude. With a maximum sequence length of $2^7$ gates and
128 repetitions of each sequence, RPE bounds the calibration at $\pm \pi/2^8$,
or a fractional uncertainty of 0.8\%. In practice, however, we observe much
lower statistical uncertainty, similar to the behavior observed in
\cite{Rudinger2017}, and repeated calibration produces the same microwave amplitude
to within 0.04\%. Following the full set of RPE trials for each type of error, we
repeat the amplitude calibration and find that it is consistent to
within 0.2\%.

We initially calibrate the microwave frequency to within 1~Hz of the qubit
resonance and find that the qubit frequency drifted less than 2~Hz after each
data set. The gate error associated with this slow frequency drift is $\approx
10^{-8}$ per gate; compared with the $\approx 10^{-6}$ per gate error associated
with drifting amplitude, this error is inconsequential. With these stable,
well-calibrated gates, we inject known errors with a significantly larger
magnitude as described in the next section.

\section{\label{sec:errors}Injecting additive errors}
\subsection{\label{sec:MeasurementError}Measurement error}
\begin{figure}
	\includegraphics[width=0.48\textwidth,]{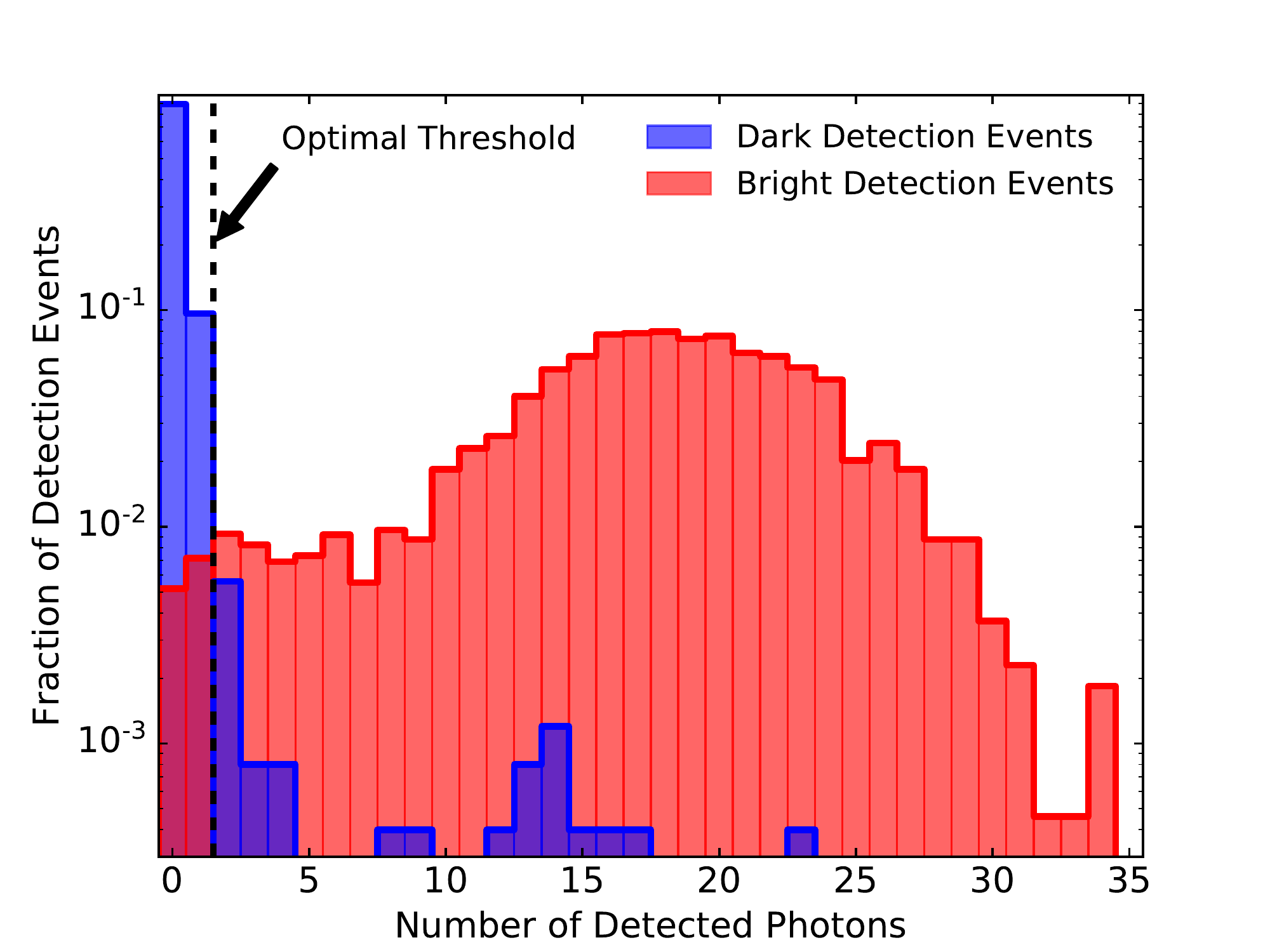}
	\caption{\label{fig:Histograms}
	Relative distribution of observed photon counts for putative $\ket{0}$
	(dark) and $\ket{1}$ (bright) states. Each experiment is repeated 2500
	times. The dashed line shows the optimal detection threshold of 2 photons.
	Larger detection errors are introduced by choosing a different threshold. At
	the optimal threshold, the measurement error is 1.2\% (sum of
	the left-most two red bars).}
\end{figure}

We inject error in the measurement process by intentionally misinterpreting the
results of our detection operation. The detection consists of counting photons
scattered from the ion over a specified duration.
When the ion is measured in the $\ket{1}$ state (bright), we collect
on average 19 photons during the 400~us detection interval, and when the ion is
measured in the $\ket{0}$ state (dark), we collect 0.1 photons.
The experimental dark and bright measurement histograms
(\reffig{fig:Histograms}) are nearly Poissonian, except for mixing of the two
Poissonians caused by errors in preparation and by the low end tail observed in
bright detection events due to off resonant scattering of the detection light
\cite{Noek2013}. The measurement error associated with these histograms (before
error injection) is 1.2\%.

To distinguish which state is present after a measurement, we define a threshold
for the number of detected photons. If the detected number of photons is less
than the chosen threshold, we infer that the ion was in the $\ket{0}$ state;
otherwise we infer that it was in the $\ket{1}$ state. The optimal threshold in
our experiment was two photons. By intentionally choosing a suboptimal
threshold, we introduced an error source in post-processing that is easy to
quantify.

In order to convert from a choice of threshold to an additive error estimate
$\delta_{\mathrm{meas}}$, we must identify the maximum error over all  possible
sequence outcomes. From the measured reference histograms, we calculate the probability that the
chosen threshold mis-identifies an observed photon count as bright (when the ion
was dark) or dark (when the ion was bright). For a general sequence, we do not
know which of the results to expect, so we take the maximum of these
probabilities as our pessimistic estimate of $\delta_{\mathrm{meas}}$.

\subsection{\label{sec:StatePrep}Preparation error}
\begin{figure}
	\includegraphics[width=0.48\textwidth,]{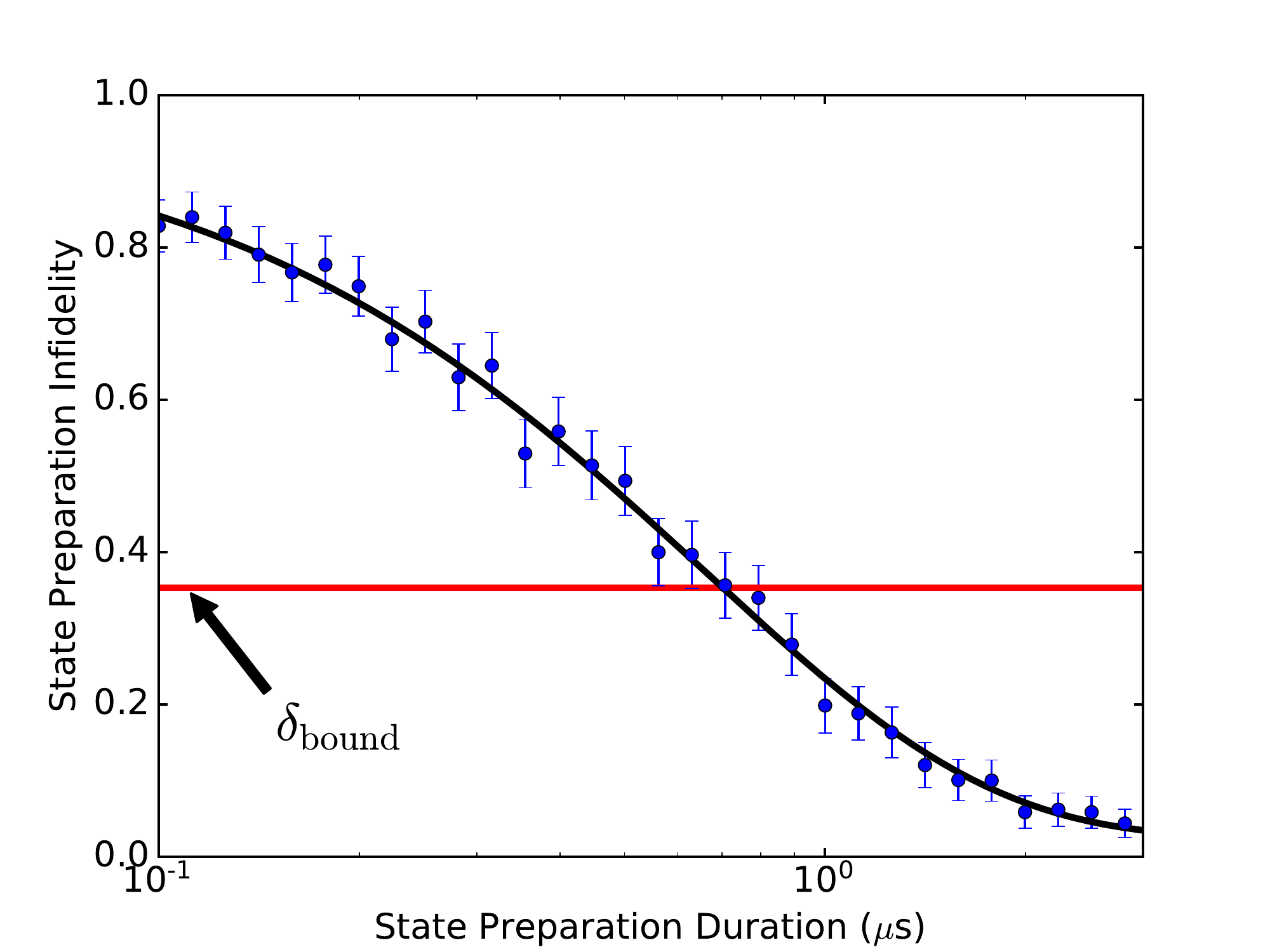}
	\caption{\label{fig:StatePrep}
	State preparation error as a function of operation duration. We
	typically use a 50 $\mu$s optical pumping duration to ensure high fidelity
	initialization in $\ket{0}$; intentionally using a shorter duration injects
	error. Over the plotted range, the exponential fit to the infidelity $E$ as
	a function of initialization duration $t$ is $E(t) = 0.95 e^{-1.5 t}+0.025$.
	At short duration, the infidelity limit is
	due to asymmetric state populations after Doppler cooling. At
	long duration, the apparent initialization infidelity is limited by
	measurement error. The $1/\sqrt{8}$ limit for additive error in RPE
	corresponds to a preparation time of about 700~ns.}
\end{figure}

The preparation of the $\ket{0}$ state is a stochastic process. The initial
state after cooling is a mixture of $^2S_{1/2}$ manifold states. During
preparation, the population in the $\ket{0}$ state exponentially approaches some
limiting value close to one within a time $\approx 2~\mu$s as depicted in
\reffig{fig:StatePrep}. In order to artificially introduce preparation errors,
we simply limit the preparation time.

At the laser frequency used for preparation, 369~nm light excites the
$^2S_{1/2}, F=1$ to $^2P_{1/2}, F^{\prime}=1$ transitions (see
\reffig{fig:experiment}a). From the $F^{\prime}=1$ states, the ion may decay
into any of the $^2S_{1/2}$ manifold states. Repetition of this scattering
process depletes population from the $^2S_{1/2}, F=1$ manifold. With our beam
polarization, we expect the rates of scatter from the $\ket{\pm1}$
states to be symmetric and faster than the scatter from the $\ket{1}$ state.
Because the $\ket{\pm1}$ states are unaffected by the gates of the RPE sequence,
population in these states after initialization is always measured in the bright
state and may or may not appear as an error, depending on the ideal result of
the sequence. In the worst case, this population sums with the population in the
$\ket{1}$ state to constitute the additive error of initialization.

In this worst case, the probability of initialization error after a preparation
interval $t$ is equal to the population in the $F=1$ manifold after that
interval. \reffig{fig:StatePrep} shows this population as a function of the
preparation time $t$ in our diagnostic experiment. We use this result to convert
between $t$ and $\delta_{\mathrm{prep}}$. For example, we observe that a
preparation time of about 700~ns corresponds to the theoretical
$\delta = 1/\sqrt{8}$ threshold for RPE success. Because this comparison
represents a worst case error, the RPE protocol is expected to tolerate even
shorter preparation times in some instances.

\subsection{\label{sec:PhaseDamping}Phase damping error}
In order to investigate an additive error that scales with the length of the
gate sequence, we illuminate the ion during the entire sequence with dim 369~nm
light tuned to the ${}^2S_{1/2}, F=1 \rightarrow {}^2P_{1/2}, F^{\prime}=0$
transition. A 369 nm photon scattering off the ion causes it to
excite and spontaneously decay if and only if the ion is in the $F=1$ manifold
(including the $\ket{1}$ state). Because we rotated the polarization of the 369
nm light to excite $\sigma^+/\sigma^-$ transitions preferentially, any
population that decayed to the $\ket{\pm 1}$ states was quickly re-excited and
transferred to the $\ket{1}$ state. The effect of this aggregate process,
considered as an error, is to dephase the qubit: it does not lead to population
mixing of the $\ket{0}$ or $\ket{1}$ states, but it does lead to decay of
superpositions of those states. Nielsen and Chuang refer to this as
\textit{phase damping} \cite{Nielsen2010}; an example of controllably inducing a
dephasing/phase damping channel is given by \citet{Urrego2018}. The decay
events we introduce through this scattering process occur discretely and
independently, so we expect the aggregate error probability they induce to
approach one as a simple exponential in time.

We control the intensity of the 369 nm light addressing the ion by changing the
RF power applied to an AOM. Calibrating the absolute magnitude
of the phase damping errors is challenging because we do not have a direct
measurement of the rate of photon scatter by the ion. To the extent that the AOM
responds proportionally to RF power, however, we can estimate the relative
scattering rates at different AOM settings. In the course of performing RPE
experiments with a known rotation angle, we can also directly observe the
additive error we introduce in each sequence by comparing results with and
without the injected errors.

Because the phase damping error only affects superpositions of $\ket{0}$ and
$\ket{1}$ states, its effective instantaneous strength changes over the course
of each RPE sequence as the superposition state changes.
As long as the characteristic time scale of the error is
longer than the $\pi$-time of the applied rotations and many rotations
are applied, it is possible to average this effect out and treat the error
as if it has a constant effective strength. 

We note as an aside that this approximation breaks down, and the error behavior
changes notably, when the 369 nm intensity is high and the inverse scattering
rate is faster than the gate time. In this case, the possibility of excitation
and decay constantly project the state and the unitary gates are suppressed.
This limit was evident in experiments performed with large injected errors,
although it is not relevant for the results presented in the next section.

\section{\label{sec:results}Results}
For all experiments testing the robustness of RPE to injected error, we used a
maximum sequence length of $2^7$ calibrated $Y_{\pi/2}$ gates and a
constant number of samples ($M_n$) independent of the sequence length. For each
trial of the RPE protocol, we computed the RPE estimate $\theta_{\mathrm{est}}$
of the gate angle using the pseudocode in the supplemental information to
\cite{Rudinger2017}. For each value of injected error, we conducted between 25
and 100 trials of the RPE protocol. The rate of failure, based on
\refeq{eq:range}, is defined as the fraction of experiments where
$\left| \theta_{\mathrm{est}}-\pi/2 \right| > \pi/2^8$.

\subsection{Detection}
\begin{figure}
	\includegraphics[width=0.48\textwidth,]{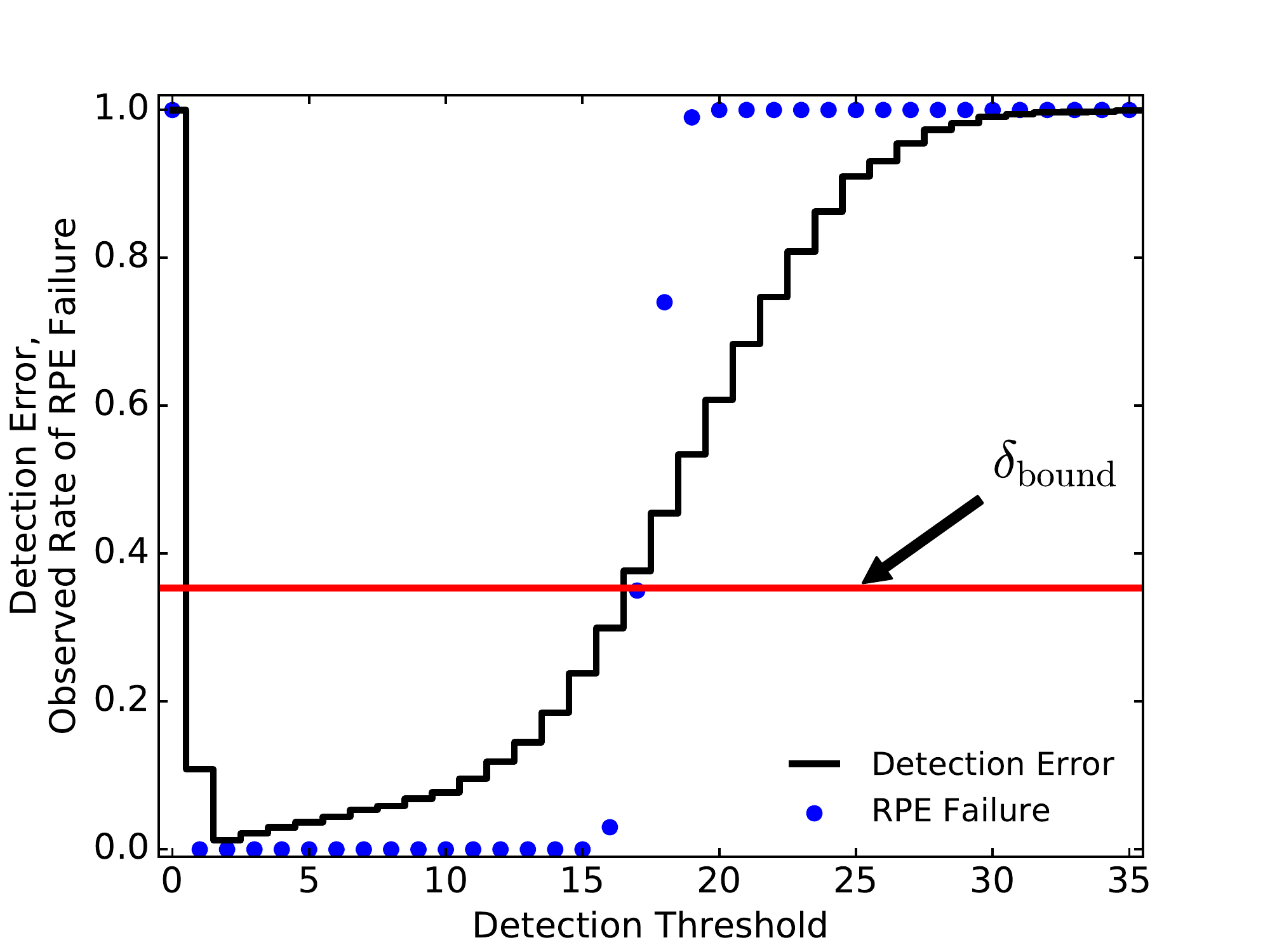}
	\caption{\label{fig:RPE-Detection}
	Observed RPE failure rate as a function of the detection threshold
	(circles). The solid black curve shows the maximum detection error
	probability, $\delta_{\mathrm{meas}}$, defined in
	\refsec{sec:MeasurementError} and computed from the histograms in
	\reffig{fig:Histograms}.  We observe a sharp onset of RPE failure when the
	threshold is set at 17 counts, in good agreement with when
	$\delta_{\mathrm{meas}} > \delta_{\mathrm{bound}}$.  Similarly, RPE also fails when
	enough dark detections are mislabeled as bright, a situation which only
	occurs for a threshold of zero.}
\end{figure}

A common error source in experimental trapped-ion measurements comes from drift
in the intensity of the detection light at the ion, due for example to positional
drift of the beam. This drift changes the photon scattering rate
and leads to a different optimal threshold as described in
\refsec{sec:MeasurementError}. In order to mimic this error and test RPE, we
intentionally degrade the measurement operation in RPE by changing the detection
threshold used in post-processing. Each RPE sequence used $M_n=32$ samples, and
the observed rate of failure for 100 trials is shown in
\reffig{fig:RPE-Detection}. RPE succeeds in accurately estimating the gate angle
despite large measurement error. As expected (see Eq. (V.16) in
\cite{Kimmel2015}), we observe a steep transition between success and failure
near the theoretical bound of $\delta_{\mathrm{meas}} \approx 1/\sqrt{8}$.

\subsection{Initialization}
We intentionally degrade the fidelity of our state preparation by shortening its
duration, with consequences described in \refsec{sec:StatePrep}. While
preparation is not frequently a large source of error in quantum applications,
it can become one when the preparation involves entangled states, such as for recent
clock and sensor experiments \cite{Ruster2017, Ockeloen2013}.
We chose to explore the combined effects both of injecting error
and of changing the number of samples ($M_n$) used in the protocol.

\begin{figure}
	\includegraphics[width=0.48\textwidth,]{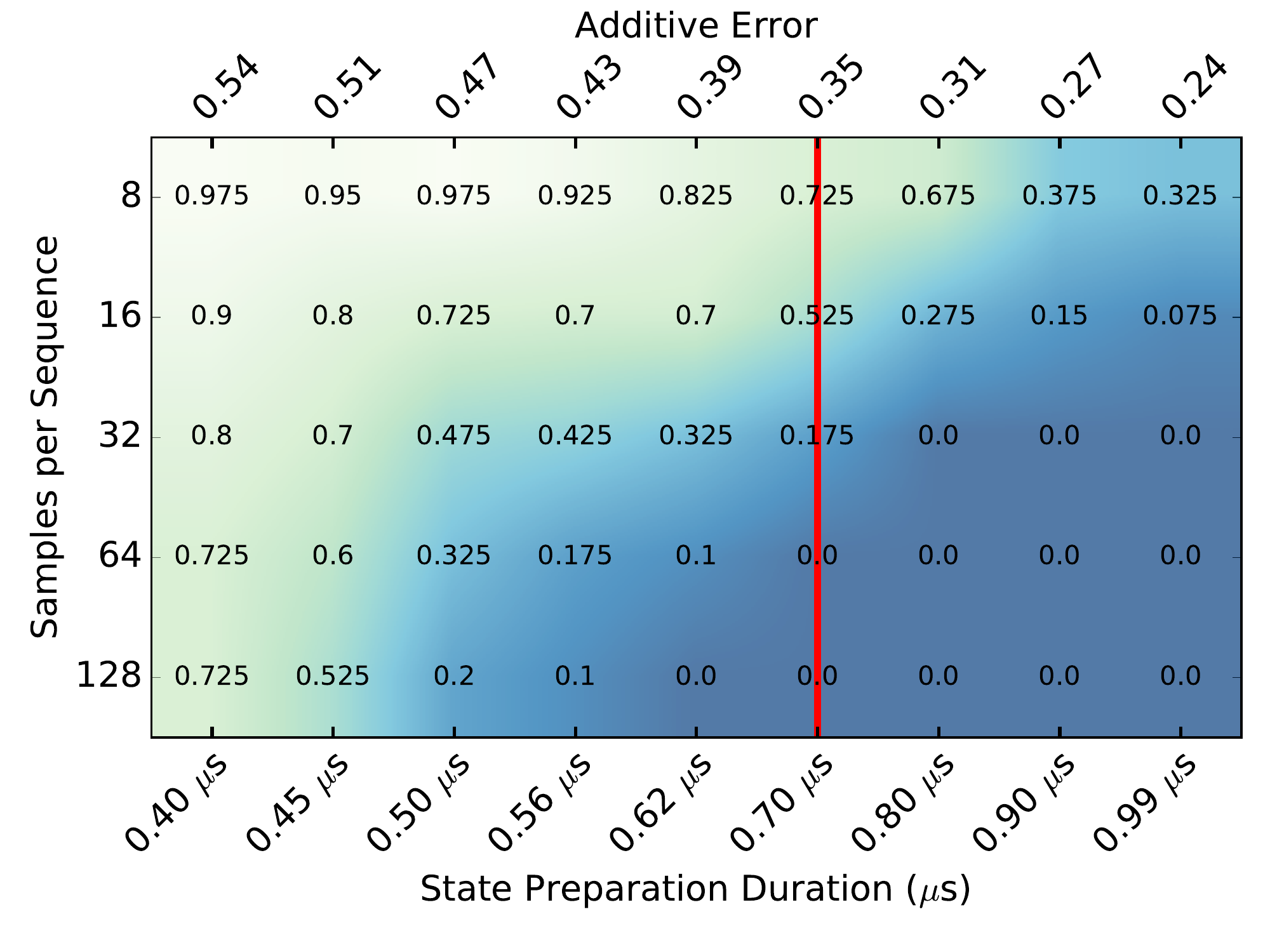}
	\caption{\label{fig:RPE-SP}
	Observed RPE failure rate as a function of qubit preparation duration
	(bottom axis) and number of samples (left axis).
	The background shading also indicates
	the observed failure rate. The top axis shows the maximum additive error
	$\delta_{\mathrm{prep}}$ corresponding to the qubit preparation duration.
	We estimate that $\delta_{\mathrm{prep}}$ exceeds $\delta_{\mathrm{bound}}=1/\sqrt{8}$ beyond 700~ns (vertical red
	line). The boundary of the onset of failure shifts as a
	function of the number of samples as expected.
	}
\end{figure}

In \reffig{fig:RPE-SP}, we show the rate of RPE success for
$\delta_{\mathrm{prep}}$ between 0.54 (0.40~$\mu$s) and 0.24 (0.99~$\mu$s).
For the center row of
\reffig{fig:RPE-SP}, we performed $M_n=32$ samples of each RPE sequence, and we
observe agreement between the onset of failure and the $1/\sqrt{8}$ bound based
on our estimate of additive error. The number of samples required to reliably
achieve a correct estimate increases with the injected error, as indicated by
the prescription for $M_n$ given by Eq. (V.17) in \cite{Kimmel2015}. For
$M_n<32$, RPE fails frequently even with injected error below the bound due to
increased measurement shot noise.

\subsection{Gate Error: Phase Damping}
Spontaneous decay and phase damping bound the coherence time of quantum
computations. These processes will bound the maximum sequence length that can be
used in the RPE protocol before the additive error exceeds its bound. In order
to observe this effect, we intentionally induce phase damping by
allowing 369 nm (measurement) light to leak into our experiment, with its
intensity reduced by $\{-20, -23, -26, -100\}$~dB from our
measurement settings. This process is described in greater depth in
\refsec{sec:PhaseDamping}.

\begin{figure}
	\includegraphics[width=0.48\textwidth,]{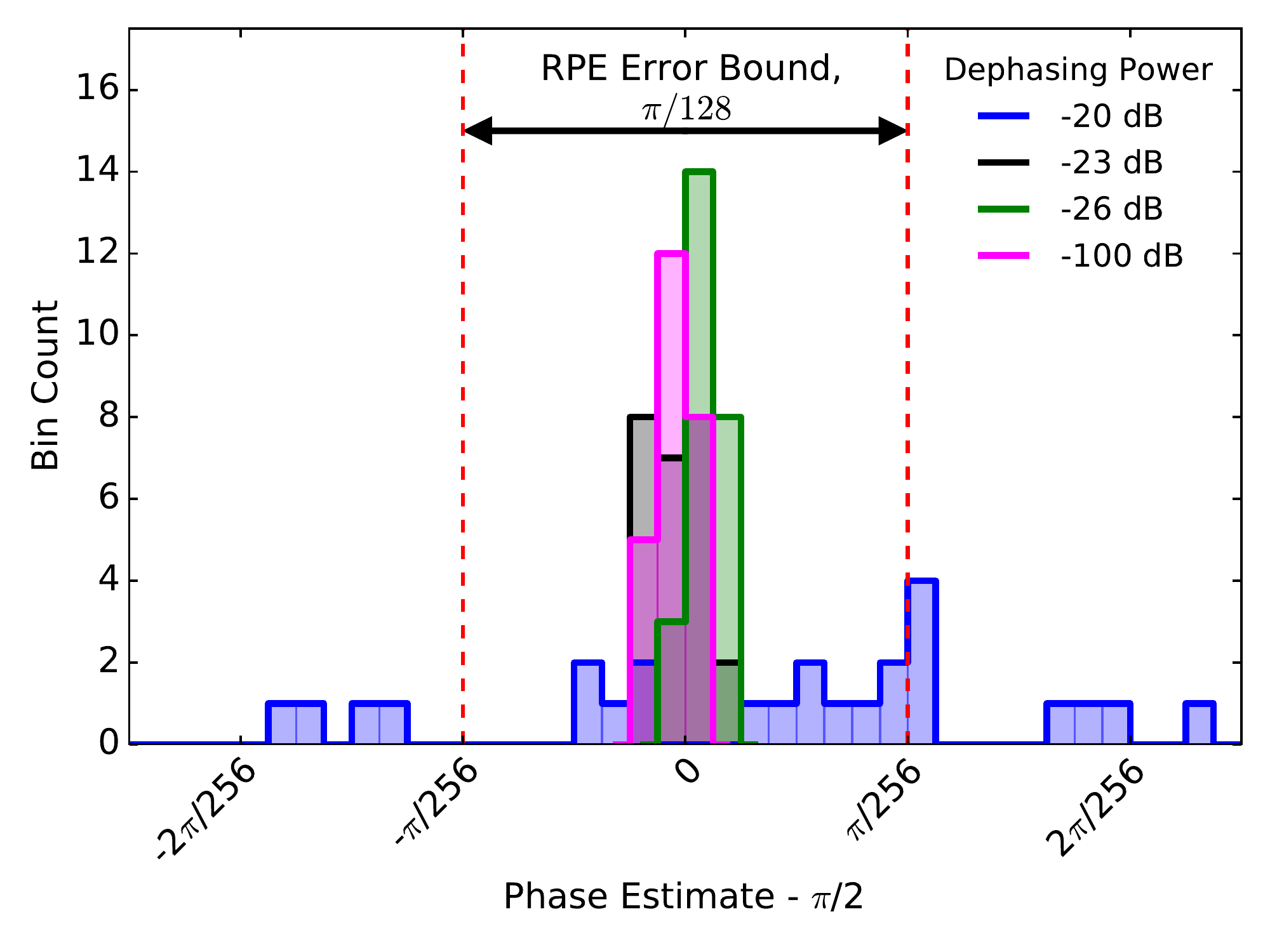}
	\caption{\label{fig:RPE-Phase-Damping}
	Histograms of rotation angle estimates from RPE with and without additional
	phase damping errors. The bounds for protocol success are shown as vertical
	dashed lines. We injected 369 nm (measurement) light at four intensities to
	drive phase damping. RPE only fails at the strongest intensity (-20 dB),
	although even here the majority of trials succeed. At lower intensities, we
	observe that the distribution of estimates is significantly tighter than the
	provable bounds reported by the protocol.
	}
\end{figure}

In \reffig{fig:RPE-Phase-Damping} we compare the histograms of RPE estimates for
these four levels of laser intensity. Because the intensities are
well below saturation, we expect spontaneous scattering to be proportional to
intensity. Although assigning an absolute phase damping error rate
is difficult in our experiment, we observe the same
qualitative behavior for RPE failure as is observed for other error sources: the
onset of RPE failure is abrupt, and RPE succeeds over a wide range of injected
phase damping errors rates.

In contrast to the detection and initialization errors described previously, the
total additive spontaneous emission error for a given gate sequence grows with
sequence length. Depending on the type of error, some sequence lengths within
the RPE protocol can be more likely to cause the RPE estimate to fail.
We observed that for the strongest scattering laser intensity tested (-20 dB),
the RPE estimate of the rotation angle only exceeded the bound of \refeq{eq:range}
at the longest sequence length leading to a clustering of
failure events within twice the claimed range of the true value. In the tests
with injected preparation or measurement errors, failures were equally likely
for any sequence length, resulting in a flatter error distribution.

\section{\label{sec:conclusion}Conclusions}
We have shown experimentally that the RPE protocol is robust to errors which
occur during state preparation, measurement, and gates, in good agreement with
the additive error bound of $1/\sqrt{8}$. Our results are
promising for noisy or initially uncalibrated experiments, where the
efficiency and robustness of RPE makes it an effective tool for automated gate
calibrations.

In order to apply RPE efficiently and successfully, we find that it is important
to take into account the dependence of the success probability on the number of
samples ($M_n$). A greater number of samples becomes more important when the
additive error is high, so, in order to maximize efficiency, it is useful to
bound the expected error.

Similarly, we observe that the failure of RPE changes in nature
depending on whether the error is constant or depends on the
sequence length. When initialization and detection errors are larger than
gate errors, failures in the RPE protocol can result in estimates that are very
far from the true value. In a quantum computing context, this could lead to rare
events of mis-calibration causing gates with poor behavior to
persist until a new calibration. This catastrophic behavior needs to be treated
differently with respect to fault tolerance than typical small-angle coherent
errors.

We expect that RPE can also find use within the quantum sensor community.  RPE
could allow a user to operate a sensor in a poorly controlled environment and
with poor initialization and detection operations, as long as the interaction
of interest can be repeated reliably. Furthermore, the robustness of RPE
to poorly calibrated operations makes it particularly appealing for automated
initial calibration of sensors. The main caveat for sensor applications appears
to be the need to prepare two orthogonal initial states, although RPE can
tolerate significant errors in this operation. In practice, this requirement
would be similar to the often-used technique of preparing alternating states for
clock measurements \cite{Brewer2019}.

Extensions of RPE to non-static parameters and to two-qubit
gates are possible directions for future research. Ion temperature and laser
intensity variations are just two examples of time-varying parameters.
Depending on the root cause, the rotation angle of a gate
may change reliably as a function of sequence length or instead may drift
randomly; it should be
possible to modify RPE to address or mitigate both of these possibilities.

An extension of RPE that calibrates two-qubit gate angle seems straightforward.
One simple and necessary adjustment is to rework RPE to account for measurements
from two qubits. However, the full calibration of two-qubit gates typically
involves more control parameters, so an important challenge is to describe a
routine based on RPE that is capable of calibrating all these parameters
together both efficiently and robustly. 

\begin{acknowledgments}
	The authors would like to thank Shelby Kimmel for fruitful discussions
	leading to the development of these experiments. Additionally, we
	acknowledge Craig Clark, Roger Brown, Kevin Schultz, Layne Churchill, and
	Kenton Brown for their careful reading of the manuscript. This work was
	funded by the IARPA STIQS study and GTRI Internal Research and Development.
\end{acknowledgments}

\bibliography{RPE}

\end{document}